\documentclass[11pt,aps,prb,preprint]{revtex4}   
\usepackage{amsmath}    
\usepackage{graphicx}   
\begin{document}
\title{Entanglement Properties and Phase Diagram of the Two-Orbital Atomic Hubbard Model}
\author{A. Avella, F. Mancini, G. Scelza}
 \affiliation{Dipartimento di Fisica ``E.R. Caianiello'' - Unit\`a CNISM
di Salerno \protect\\ Universit\`a degli Studi di Salerno, I-84081
Baronissi (SA), Italy}
\author{S. Chaturvedi} \affiliation{School of Physics, University
of Hyderabad, Hyderabad 500 134, India}

\begin{abstract}
We study the two-orbital Hubbard model in the limit of vanishing
kinetic energy. The phase diagram in the $V-J$ plane, with $V$ and
$J$ denoting the interorbital hybridization and exchange coupling
respectively, at half filling is obtained. A
singlet(dimer)-triplet transition is found for a critical value of
the ratio $V/J.$ The entropy of formation, both in the mode and in
the particle picture, presents a jump as the same critical line in
conformity with the suggested relation between criticality and
entanglement.
\end{abstract}
 \maketitle
\section{Introduction}
The two-orbital Hubbard model has recently come into limelight as
a minimal model capable of describing the phenomenon of orbital
selective Mott transition experimentally observed in certain
materials \cite{Anisimov_02,Medici_05,Ferrero_05}. This was
demonstrated \cite{Scelza} through a simple two-pole approximation
within the framework of the composite operator method
\cite{Mancini_04}. In the present work, we report a preliminary
aimed at improving the two-pole approximation by using a new
basis, the one that solves the system when reduced to a single
atom.
\section{The model}
The Hamiltonian describing the two-orbital Hubbard model in the
limit of vanishing kinetic energy (i.e., reduced to a single atom)
read as
\begin{equation}\label{eqn:ham}
  H=V\sum_{\alpha\neq\beta}c^{\dag}_{\alpha}c_{\beta}-\mu\sum_{\alpha}c^{\dag}_{\alpha}c_{\alpha}+
  U\sum_{\alpha}D_{\alpha}+U'n_1n_2-\frac{1}{2}Jn_{1\mu}n_{2\mu}+
  J\sum_{\alpha\neq\beta}c_{\alpha\uparrow}c_{\alpha\downarrow}c^{\dag}_{\beta\downarrow}c^{\dag}_{\beta\uparrow},
\end{equation}
where
$c^{\dag}_{\alpha}=(c^{\dag}_{\alpha\uparrow},c^{\dag}_{\alpha\downarrow})$
is the electronic creation operator in spinorial notation in the
orbital $\alpha,$ $D_{\alpha}$ is the double occupancy operator in
the orbital $\alpha,$ $n_{\alpha\mu}$ is the ($\mu=0$ or
$n_{\alpha}$) charge and ($\mu=1,2,3$) spin density operator in
the orbital $\alpha,$ $V$ is the interorbital hybridization, $\mu$
is the chemical potential, $U$ is the intraorbital Coulomb
repulsion, $U'$ is the interorbital Coulomb repulsion, $J$ is the
exchange interorbital interaction. Hereafter, we will use $U$ as
the unit of energy and we will fix, as usual, $U'=U-2J.$
\section{Phase diagram and entanglement}
At zero temperature and half filling ($N=2$), by studying the
exact solution in terms of eigenvalues and eigenvectors of $H,$ it
is possible to show that the system undergoes a phase transition
between a singlet (diner) state
$[|\uparrow;\downarrow\rangle\otimes|\uparrow\downarrow;0\rangle]$
and a triplet one $[|\uparrow;\uparrow\rangle]$ at a critical
value of the interorbital hybridization: $V_c=\sqrt{2}J.$

The eigenvaues and eigenvectors in the half-filling sector read as
\begin{align}
  |1\rangle &=\frac{1}{\sqrt{2}}(|\uparrow\downarrow;0\rangle+|0;\uparrow\downarrow\rangle),\\
  |2\rangle &=\frac{1}{\sqrt{2(a^2+1)}}[a(|\uparrow;\downarrow\rangle-|\downarrow;\uparrow\rangle)+
  |\uparrow\downarrow;0\rangle-|0;\uparrow\downarrow\rangle],\\
  |3\rangle &=\frac{1}{\sqrt{2(b^2+1)}}[b(|\uparrow;\downarrow\rangle-|\downarrow;\uparrow\rangle)+
  |\uparrow\downarrow;0\rangle-|0;\uparrow\downarrow\rangle],\\
  |4\rangle &=|\uparrow;\uparrow\rangle,\\
  |5\rangle &=\frac{1}{\sqrt{2}}(|\uparrow;\downarrow\rangle+|\downarrow;\uparrow\rangle),\\
  |6\rangle &=|\downarrow;\downarrow\rangle,\\
  E_1 &=-2\mu+2V+U-J,\\
  E_2&=-2\mu+2V+\frac{1}{2}(U+U')+J-\frac{1}{2}\sqrt{(U-U')^2+16V^2},\\
  E_3&=-2\mu+2V+\frac{1}{2}(U+U')+J+\frac{1}{2}\sqrt{(U-U')^2+16V^2},\\
  E_4&=-2\mu+2V+U'-J,\\
  E_5&=E_4,\\
  E_6&=E_4,
\end{align}
where
\begin{align}
  a&=-\frac{1}{4V}\biggl(U-U'+\sqrt{(U-U')^2+16V^2}\biggr),\\
  b&=-\frac{1}{4V}\biggl(U-U'-\sqrt{(U-U')^2+16V^2}\biggr).
\end{align}
In this system, it is also possible to study both the particle
entropy \cite{Eckert} and the mode entropy \cite{Zanardi}. The
particle entropy $S_p$ requires the calculation of the concurrence
$C$:
\begin{equation}\label{eqn:essep}
  S_p=-\frac{1+\sqrt{1-C^2}}{2}\log_2\biggl(\frac{1+\sqrt{1-C^2}}{2}\biggr)-
    \frac{1-\sqrt{1-C^2}}{2}\log_2\biggl(\frac{1-\sqrt{1-C^2}}{2}\biggr),
\end{equation}
\begin{equation}\label{eqn:conc}
  C=max\{0,\lambda_1-\lambda_2-\lambda_3-\lambda_4-\lambda_5-\lambda_6\},
\end{equation}
where $\{\lambda_i\}$ stands for the square roots of the
eigenvalues, taken in descending order of magnitude, of the matrix
$\rho D\rho D^{-1}.$ $\rho=\frac{e^{-\beta H}}{Tr(e^{-\beta H})}$
is the density matrix of the system ($\beta$ is the inverse
temperature) and $D=-U_{ph}\mathcal{K}$ is the dualization
operator obtained by composing the particle-hole transformation
$U_{ph}$ with the conjugation operator $\mathcal{K}.$ The mode
entropy $S_m,$ on the other hand, requires calculation of the
reduced density matrix $\rho_\beta,$ with respect to a chosen
orbital $\alpha$ ($\beta\neq \alpha$):
\begin{equation}\label{eqn:dens}
  \rho_{\beta}=\sum_i\langle i_{\alpha}|\rho|i_{\alpha}\rangle ,
\end{equation}
where $\{|i_{\alpha}\rangle\}$ stands for a complete basis set for
the orbital $\alpha.$ Then, we simply have
$$S_m=-Tr(\rho_{\beta}\log\rho_{\beta}).$$
\begin{figure}[!hbp]
\centering\includegraphics*[width=0.7\textwidth]{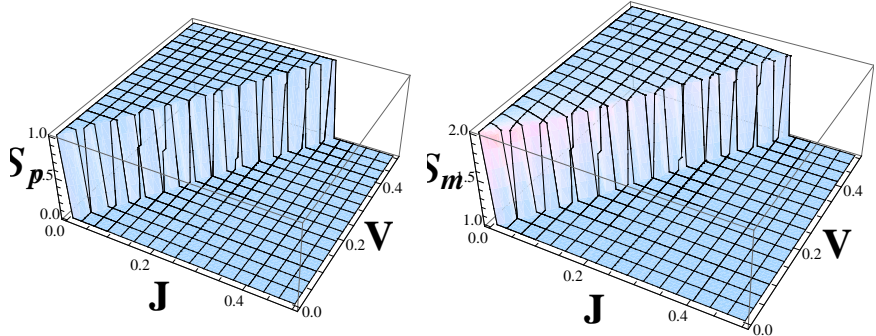}
\caption{(left) Particle entropy and (right) mode entropy at $N=2$
and $T=0$ as functions of the interorbital hybridization $V$ and
of the exchange interorbital interaction $J.$}\label{figura1}
\end{figure}
In Fig. 1, the particle entropy (left part) and the mode entropy
(right part) are reported at $N=2\mbox{ and }T=0$ as functions of
the interorbital hybridization $V$ and of the exchange
interorbital interaction $J.$ Both types of entropy show a well
defined jump exactly on the line $V=\sqrt{2}J$ where the phase
transition occurs. However, it is worth noticing that the mode
entropy, in contrast to the the particle entropy, is not capable
of discriminating between a \textit{genuine} entanglement between
substantially different elementary states (a dimer
$|\uparrow\downarrow;0\rangle$ and a singlet
$|\uparrow;\downarrow\rangle$) and the \textit{trivial}
entanglement between states arising from symmetry requirements
(the three sates of a triplet $|\uparrow;\uparrow\rangle$). As a
matter of fact, the particle entropy is the only measure correctly
accounting for an absolute lack of entanglement in the latter
case.
\section{Conclusions}
In conclusion, we have shown that both entanglement measures known
in the literature (particle entropy and mode entropy) are capable
of capturing the essential physics of the atomic two-orbital
Hubbard model. In particular, their jumps can be used for
determining the location, in the phase diagram, of the transition
line separating the singlet (dimer) state and the triplet.


\begin{thebibliography}{00}
\bibitem{Anisimov_02}
V.~I. Anisimov, I.~A. Nekrasov, D.~E. Kondakov, T.~M. Rice, and M.
Sigist,
  Eur.~Phys.~J.~B {\bf 25},  191  (2002).
\bibitem{Medici_05}
L. de~Medici, A. Georges, and S. Biermann, Phys.~Rev.~B {\bf 72},
205124
  (2005).
\bibitem{Ferrero_05}
M. Ferrero, F. Becca, M. Fabrizio, and M. Capone, Phys.~Rev.~B
{\bf 72},
  205126  (2005).
\bibitem{Scelza} A. Avella, F. Mancini, S. Odashima, G. Scelza, \textit{Physica
C}\textbf{460-462}, 1068 (2007).
\bibitem{Mancini_04}
F. Mancini and A. Avella, Adv.~Phys. {\bf 53},  537  (2004).
\bibitem{Eckert} K. Eckert, J. Schliemann, D. Bru\ss, M.
Lewenstein, \textit{Ann. Phys.} \textbf{299}, 88 (2002).
\bibitem{Zanardi} P. Zanardi, \textit{Phys. Rev. A} \textbf{65},
042101 (2002).
\end{thebibliography}
\end{document}